\documentclass[floatfix,eqsecnum,aps,nofootinbib,showpacs]{revtex4-1}

\usepackage{amsfonts, amssymb, amscd, latexsym}
\usepackage{graphicx}
\usepackage[latin1]{inputenc}
\usepackage{amssymb}
\usepackage{amscd}
\usepackage{mathrsfs}
\usepackage{mathbbol}
\usepackage{float}
\usepackage[usenames,dvipsnames]{color}
\usepackage{verbatim}
\usepackage{graphicx,epstopdf}
\usepackage[dvips]{epsfig}
\usepackage{amsmath}
\usepackage{amssymb}
\usepackage{amscd}
\usepackage{mathrsfs}
\usepackage{graphicx}
\usepackage{mathbbol}
\usepackage{float}
\usepackage[usenames,dvipsnames]{color}
\usepackage{amssymb}
\usepackage{graphicx}

\newtheorem{theorem}{Theorem}[section]

\newcommand{\ket}[1]{\left | #1 \right\rangle}
\newcommand{\bra}[1]{\left \langle #1 \right |}

\renewcommand{\epsilon}{\varepsilon}

\def\bdyB1{S^1}

\newcommand{\Guone}{\ensuremath{\mathrm{U}(1)}}
\newcommand{\Gsuthree}{\ensuremath{\mathrm{SU}(3)}}
\newcommand{\Gsutwo}{\ensuremath{\mathrm{SU}(2)}}
\newcommand{\Gsufive}{\ensuremath{\mathrm{SU}(5)}}

\newcommand{\Gsoten}{\ensuremath{\mathrm{SO}(10)}}
\newcommand{\Gesix}{\ensuremath{\mathrm{E}_6}}

\begin{document}
\preprint{NIKHEF/2012-001}

\title{Symmetry breaking, subgroup embeddings and the Weyl group}

\author{Damien P. George}\email{dpg39@cam.ac.uk}
\affiliation{Nikhef Theory Group, Science Park 105,
1098 XG Amsterdam, The Netherlands}
\author{Arun Ram}\email{aram@unimelb.edu.au}
\affiliation{Department of Mathematics and Statistics, The University of Melbourne, Victoria 3010, Australia}
\author{Jayne E. Thompson}\email{thompson.jayne2@gmail.com}
\affiliation{ARC Centre of Excellence for Particle Physics at the Terascale, School of Physics, The University of Melbourne, Victoria 3010, Australia}
\author{Raymond R. Volkas}\email{raymondv@unimelb.edu.au}
\affiliation{ARC Centre of Excellence for Particle Physics at the Terascale, School of Physics, The University of Melbourne, Victoria 3010, Australia}

\begin{abstract}
We present a systematic approach to finding Higgs vacuum expectation values, which break  a symmetry $G$ to differently embedded isomorphic copies of a subgroup $H \subset G$ . We give an explicit formula for recovering each point in the vacuum manifold of a Higgs field which breaks $G\rightarrow H$. In particular we systematically identify the vacuum manifold $G/H$ with linear combinations of the vacuum expectation values breaking $G \rightarrow H_1 \rightarrow \dots \rightarrow H_{\it l}$. We focus on the most applicable case for current work on grand unified theories in extra dimensional models and low-energy effective theories for quantum chromodynamics. Here the subgroup, $H$, stabilizes the highest weight of the fundamental representation leading to a simple expression for each element of the vacuum manifold; especially for an adjoint Higgs field. These results are illustrated explicitly for adjoint Higgs fields and clearly linked to the mathematical formalism of Weyl groups. We use the final section to explicitly demonstrate how our work contributes to two contemporary high-energy physics research areas.
\end{abstract}


\maketitle


\section{Introduction}

Symmetry breaking is a crucial aspect of modern particle physics. In particular the symmetry breaking sectors of theories extending the standard model are studied intensively. Many of the most puzzling problems in generic standard model extensions, such as the gauge hierarchy and parameter proliferation problems, arise because of the use of elementary scalar fields to spontaneously break symmetries.  Deeper insights into both the physics and mathematics of symmetry breaking are thus worth having.

%



The majority of model building scenarios consider a gauge symmetry $G$ which is spontaneously broken to a subgroup $H$.  Of special interest here are models where $G$ is broken simultaneously to several isomorphic but differently embedded subgroups $H$.  Below we enumerate several general classes of models where this mechanism is apparent.  In such models, each isomorphically embedded subgroup is given by $gHg^{-1}$ for some $g \in G$, and is identified with a point in the vacuum manifold $G/H$.  For a Higgs field in a $G$ representation and an associated basis of weights for this representation, we develop explicit mathematical formulas for writing each state in the vacuum manifold as a linear combination of the weights.
To make our result accessible to the model-building audience at large, we provide a careful review of the necessary mathematical tools which belong to the discipline of representation theory.  We shall also explain all our mathematical results in the physical context of spontaneous symmetry breaking.

In the adjoint representation the
weights are the roots, and there exists a basis of weights for this adjoint representation which are known as the simple roots. These can be pulled back to define a Cartan subalgebra $h_1,\dots, h_{\it l}$, consisting of adjoint Higgs vacuum expectation values (vevs) which cause the symmetry to break along $G \rightarrow H_1 \rightarrow H_2 \rightarrow \dots \rightarrow H_{\it l}$. Our formalism identifies
the vacuum manifold $G/H$ which belong to the vector
space spanned by the Cartan subalgebra. The elements of the vacuum manifold are related by a Weyl group symmetry. Given an adjoint Higgs vacuum
expectation value, $h$, breaking
  $G \rightarrow H$, a full complement of vevs breaking $G$ to different
 Cartan preserving embeddings of the subgroup $H$ can be
  obtained through this method. We give an explicit formula
   for recovering each vev.

 In our treatment, we shall concentrate on a specific and relevant case for high-energy physics model building scenarios, where $H$
  stabilizes the highest weight of the lowest
 dimensional fundamental representation. This case in fact admits the simplest
  formula for recovering all vevs breaking $G \rightarrow H$, when the
  Cartan subalgebra for $H$ is a subset of the Cartan subalgebra for $G$.

The results presented here may provide direct solutions to problems in high-energy physics research. This includes problems arising in:
\begin{itemize}
\item[i] Grand unified theories (GUTs), where so-called ``flipped'' models arise whenever there are alternative embeddings of a given GUT inside a larger GUT \cite{SMBarr1982,Derendinger1984}.
\item[ii]{Domain-wall brane scenarios which use the ``clash of symmetries'' mechanism \cite{Davidson:2002eu,Shin:2003xy,Davidson:2007cf}.  This idea was the main motivation for us to pursue the present study.\label{item:domain-wallbranes}}
\item[iii]{ The low-energy limit of Yang-Mills theory~\cite{Cho:2007ja}.\label{item:yang-mills}}
\item[iv] Whenever there are multiple copies $\Phi_1,\Phi_2,\ldots$ of a given representation of Higgs fields, with vevs $\langle \Phi_1 \rangle,\langle \Phi_2 \rangle ,\ldots$ breaking the gauge group to isomorphic but differently embedded subgroups. This is a special case of what is generally termed ``vacuum alignment''.
\end{itemize}
Each of these physical contexts is reviewed in more detail in the next section.  To show explicitly how our results can be utilized we apply them to two model building examples from the literature, Ref.~\cite{Davidson:2007cf} and Ref.~\cite{Cho:2007ja}, corresponding to contexts (ii) and (iii) listed above.
It would also be possible to apply these techniques to help identify standard model particles embedded in representations of a grand-unified gauge group \cite{Chaichian:1980jg, Slansky:1981yr}, and to help classify and construct different chains of embeddings in the context of grand unification \cite{Chaichian:1981uy}.
We have kept our analysis general and there may well be other applications for different embeddings of isomorphic subgroups.

 We focus on presenting our results in a self contained and accessible manner. We include examples of how our results relate to problems in the high-energy physics literature and explicitly apply the techniques developed here to the field. We are keen to ensure the dialogue is bicultural and accessible to mathematicians as well as physicists.

Follow the introduction, we begin in Sec.~\ref{sec:motivation} by providing a thorough physical motivation by discussing four model building scenarios. Sec.~\ref{sec:notation} gives the necessary notation, and in Secs. \ref{sec:weylgroups}-\ref{sec:embeddings} we motivate and explain the representation theory needed to understand the proof of the main result.
The proof itself appears in Sec.~\ref{sec:proof}, where we clearly state the formula for recovering the adjoint Higgs vevs which break $G$ to different embeddings of a  subgroup $H$ as linear combinations of vevs breaking $G$ along the chain $G \supset H_1 \supset H_2 \supset \dots \supset H_{\it l}$. We also treat the relation between the weights of vevs causing $G$ to break to different embeddings of a subgroup $H$, for a non-adjoint Higgs field.
In Sec.~\ref{sec:applyresults} we explicitly apply the main result to two concrete examples from the contemporary physics literature, thus placing our results in the context of model building scenarios.  We conclude in Sec.~\ref{sec:conclusion}.

\section{Motivation}\label{sec:motivation}

We now explain some of the physical contexts for our work in more detail.

\subsection{Flipped grand unification}
\label{sec:flippedgrandunified}
The simplest example of flipped grand unification is flipped $\Gsufive$  \cite{SMBarr1982,Derendinger1984}. The relationship between standard and flipped $\Gsufive$ may be explained using two different embeddings of $\Gsufive \times \Guone$ inside $\Gsoten$.  Call these two subgroups $\Gsufive_{s} \times \Guone_{X_s}$ and $\Gsufive_{f} \times \Guone_{X_f}$.  One of these embeddings has been labeled $s$ for ``standard'', and the other $f$ for ``flipped''.  The selection of one as standard is purely a matter of convention; the important issue is the relationship between the two embeddings.  Having decided to call one embedding ``standard'', the standard weak hypercharge generator is identified as the $Y_s$ obtained through $\Gsufive_s \to \Gsuthree \times \Gsutwo \times \Guone_{Y_s}$. By contrast in the flipped case, the weak hypercharge generator is $Y_f$, which arises from a second embedding of SU(5) inside SO(10); namely $\Gsoten \to \Gsufive_f \times \Guone_{X_f} \to [\Gsuthree \times \Gsutwo \times \Guone_{Y_f}] \times \Guone_{X_f}$, where $X_f$ is a linear combination of $Y_s$ and $X_s$. $\Guone_{Y_f}$ is not a subgroup of $\Gsufive_s$, in fact $Y_f$ is a linear combination of $Y_s$ and $X_s$ which is linearly independent of $X_f$.

This concept can be extended through $\Gesix$ grand unification.  The subgroup chain
\begin{equation}
\Gesix \to \Gsoten \times \Guone'' \to \Gsufive \times \Guone' \times \Guone''
\end{equation}
can be shown to contain three possible candidates for weak hypercharge: standard, flipped, and double-flipped.  Standard hypercharge is a generator of $\Gsufive$.  The flipped choice is a linear combination of standard hypercharge and the $\Guone'$ generator, while the double-flipped choice also involves an admixture of the generator of $\Guone''$.  Once again, each of these candidate hypercharges is actually a generator of a differently-embedded $\Gsufive$ subgroup within $\Gesix$.

\subsection{Domain-wall brane models}
\label{domain-wallbranemodels}
This work was primarily motivated by a study of domain-wall topological defects created by an adjoint scalar field ${\cal X}$ \cite{Vachaspati:2002gu}. In particular we study the case where the Lagrangian is invariant under a discrete symmetry, $Z$, and a continuous internal symmetry, $G$, but along two distinct antipodal directions the asymptotic configuration of the scalar field breaks $Z \times G$ down to differently embedded isomorphic copies of $H\subset G$. This construction has a natural manifestation in grand-unified models with gauge group $G$ and a single infinite extra dimension. Here the adjoint scalar field interpolates between two vacuum configurations preserving subgroups $H$ and $z gHg^{-1}$ (for some $z \in Z$ and $g \in G$) as a function of the extra dimensional co-ordinate\footnote{The role of the discrete symmetry breaking is to provide disconnected vacua which then serve as the boundary conditions for topologically non-trivial domain-wall solutions \cite{Dzhunushaliev:2009va,Carroll:1997pz}.  Cosmologically, one expects domain-wall formation when causally disconnected patches of spacetime acquire different vacuum configurations.}, $y$ .   The case $g = 1$ defines what may be called the standard domain wall or kink  \cite{Vachaspati:1997rr}.  In this case, the spontaneous symmetry breaking produces exactly the same unbroken subgroup $H$ on opposite sides of the domain wall.  At generic values of $y$, the configuration is also stabilized by exactly that same $H$, except for a finite number of points where the unbroken subgroup may be instantaneously larger than $H$.  The interesting fact is that for certain $g \neq 1$, domain-wall solutions can also exist.  This situation has been termed the ``clash of symmetries (CoS)'', because now the unbroken subgroups in the ``bulk'' on opposite sides of the domain wall are no longer identical, though they are isomorphic \cite{Davidson:2002eu,Shin:2003xy,Davidson:2007cf}.

One reason to be interested in CoS domain walls is the dynamical localization of massless gauge fields to the domain wall, thus effecting a dimensional reduction from a $d+1$-dimensional gauge theory to a $(d-1)+1$-dimensional gauge theory.  The idea, which is an elaboration of an original proposal due to Dvali and Shifman  \cite{Dvali:1996xe}, is as follows.  We suppose that the non-Abelian factors in the $H$ and $gHg^{-1}$ gauge theories produced on opposite sides of the wall are in confinement phase.  The underlying mechanism for this might, for example, be dual superconductivity. On the wall, the unbroken subgroup is $H\cap gHg^{-1}$, which is a subgroup of both $H$ and $gHg^{-1}$.  The idea is that the gauge fields of a certain subgroup of $H\cap gHg^{-1}$ are dynamically localized, due to the mass gap created by the confining dynamics in the bulk.  An example of this situation has been provided in Ref.~\cite{Davidson:2007cf}.  Here, $\Gesix$ breaks to differently embedded $\Gsoten \times \Guone$ subgroups in the bulk on opposite sides of the domain wall.  For appropriately chosen pairs of these subgroups, their intersection is $\Gsufive \times \Guone \times \Guone$.  By hypothesizing that the $\Gsoten$ gauge forces lead to confinement, the conclusion is that the $\Gsufive$ gauge fields should be dynamically localized on the wall\footnote{It has not been definitely established that the Dvali-Shifman mechanism works, but the heuristics are compelling.  Note that for $d>3$, the bulk dynamics is governed by a non-renormalisable gauge theory that must be implicitly defined with an ultraviolet cut-off, beyond which new physics must be invoked to complete the dynamics.  Studies of Yang-Mills theory in $4+1$ dimensions at finite lattice spacing, which acts as an ultraviolet cut-off, support the existence of a confinement phase when the gauge coupling constant is above a critical value \cite{Creutz:1979dw}.}.  This is interesting for model building when $d=4$, because the dynamically-localized $d=3$ $\Gsufive$ gauge theory could form the basis for a phenomenologically-realistic standard model extension. It is simultaneously possible to localize fermions to a domain-wall brane \cite{Visser:1985qm,Rubakov:1983bb}, thereby providing all the components for a $3+1$-dimensional grand-unified theory.

To implement the CoS mechanism we must solve the Euler-Lagrange equations for ${\cal X}$ for boundary conditions as $y\rightarrow \pm \infty$ breaking $G \times Z$ to $H$ and $z g Hg^{-1}$, respectively. Therefore it is necessary to understand how the boundary conditions breaking $G$ to $gHg^{-1}$ can be written as a linear combination of the adjoint scalar field vevs breaking $G$ along the $H_{1,2,3,\dots, {\it l}}$ branching direction in the Cartan subalgebra.

Solutions to the Euler-Lagrange equations satisfying different boundary conditions have different energies. A boundary condition preserving a symmetry $H$ can be continuously transformed into a boundary condition preserving any other isomorphic subgroup $g H g^{-1}$ inside $G$, and for some choices of $g$ solutions interpolating between the $H$- and $z g H g^{-1}$-preserving boundary conditions exist. The phenomenology of each of these domain-wall solutions is different because each different non-isomorphic intersection $H \cap g H g^{-1}$ will give rise to a different gauge theory localized on the domain wall.  Hence an exhaustive search  for the lowest energy stable domain-wall configuration must be executed. This search must range through all solutions to the Euler-Lagrange equations with different boundary conditions. In this case a systematic method for finding all the different possible configurations must be established.  To trap a copy of the standard model gauge group on the domain wall, the grand-unified gauge group must have a comparatively high rank, for example $\Gesix$ as in \cite{Davidson:2007cf}. For high rank groups a method for writing one set of boundary conditions in terms of another becomes critical.

 To find the vev for the adjoint ${\cal X}$ breaking $G$ to a subgroup $gHg^{-1}$ as a linear combination of vevs along the $H_{1,2,3,\dots , {\it l}}$ branching direction in the Cartan subalgebra, the authors of \cite{Davidson:2007cf} wrote down the Casimir operators (invariants) for a general linear combination of the Cartan subalgebra, $h^1,\dots,h^{\it l}$. The coordinates in the Cartan subalgebra space which extremize the Casimir operators correspond to linear combinations which break $G$ to certain subgroups, including $H$ and $gHg^{-1}$. The physical reason for this is: invariance of the action under the internal symmetry forces the potential to be a polynomial in the Casimir invariants. Therefore extrema of the Casimir operators correspond to degenerate minima in the vacuum manifold associated with spontaneous breaking of the internal symmetry $G$ to various subgroups, including to differently embedded isomorphic copies of a subgroup $H = H' \times {\rm U}(1)_H$. Hence the coefficients in the linear combination which extremize the Casimir invariants are precisely the components of the adjoint Higgs field in the original Cartan subalgebra basis which combine to give the ${\rm U}(1)_{gHg^{-1}}$ generator which spontaneously condenses to break $G \rightarrow gHg^{-1}$. This approach is labor intensive, and the techniques to be explained in this paper will improve upon it.

\subsection{Low-energy limit of Yang-Mills theory}
\label{sec:lowenergyyangmills}

We have found a natural motivation for our work in domain-wall formation due to the breaking of a global symmetry on cosmological scales. At the other end of the spectrum, in low energy effective models for SU(3) (and SU(2)) pure Yang-Mills gauge theories, domain walls form due to a breakdown of Weyl group symmetry caused by gluon condensation. This gives rise to a trapping of gauge fields on the domain wall. Galilo and Nedelko \cite{Galilo:2010fn} work with an effective potential generated by loop order corrections in a low energy effective field theory approach to QCD:
\begin{equation}
U_{\rm eff}=\frac{1}{12} {\rm Tr}\left(C_1\hat{F}^2 + \frac{4}{3}C_2\hat{F}^4 - \frac{16}{9}C_3\hat{F}^6\right),
\end{equation}
\noindent where the potential is confining provided $C_1>0, \ C_2>0, \ C_3 > 0$, and the non-Abelian gauge field strength tensor, $\hat{F}_{\mu\nu}$, can be written in terms of the SU(3) Lie algebra structure constants $f^{abc}$ as,
\begin{eqnarray}\label{eq:fieldstrength}
F^a_{\mu\nu} &=& \partial_\mu G^a_\nu - \partial_\nu G^a_\mu - i f^{abc} G^b_\mu G^{c}_\nu, \notag \\
\left(\hat{F}_{\mu\nu}\right)_{ bc} &=& F^a_{\mu\nu} {T^a}_{bc},\ \ \ T^a_{bc} = -i f^{abc}.
\end{eqnarray}

 The second order Casimir invariant ${\rm Tr}(\hat{F})^2 = - 3 F^a_{\mu \nu} F^a_{\mu\nu} \leq 0$, causing the minimum of the effective potential to occur at a nonzero gluon field strength.

\begin{equation}
  F^a_{\mu\nu} F^a_{\mu\nu} = \frac{4}{9 C_3^2} \left(\sqrt{C_2^2+3C_1C_3}-C_2\right)^2 \Lambda^4 > 0,
\end{equation}

\noindent where $\Lambda$ is the QCD confinement scale.

Galilo and Nedelko  \cite{Galilo:2010fn}  look at the effective potential for  $\hat{F}_{\mu\nu}= h^{\chi} B^{\chi}_{\mu\nu}$, which involves restricting the full SU(3) gauge theory to the ${\rm U}(1) \times {\rm U}(1)$ Abelian subspace, where the generators are given as a linear combination of the diagonal Gell-Mann matrices,
\begin{equation}
h^{\chi} = \chi^1 \lambda_3 + \chi^2 \lambda_8,
\end{equation}
\noindent and the associated field strength $B^{\chi}_{\mu,\nu}$ can be found by using the Abelian subalgebra version of (\ref{eq:fieldstrength}) on $B^{\chi}_{\mu} = \chi^1 G^3_{\mu} + \chi^2 G^8_{\mu}$. The minima of the effective potential are located at:

\begin{equation}
\chi = ({\rm Cos } \frac{(2n + 1)\pi}{6},  {\rm Sin } \frac{(2n + 1)\pi}{6}) \textrm{ for } n \in \{0,\dots,5\}.\end{equation}

\noindent They are related by a discrete Weyl group symmetry. The requirement that QCD remains unbroken despite a nonzero background field strength means the background field must be the average of an ensemble of gauge field configurations with a high degree of disorder and spatial variation of the direction $\chi$ in color-space. This causes different vacua to be picked out in different spatial regions. Galilo and Nedelko  \cite{Galilo:2010fn} explain that domain-wall configurations are formed by gauge fields interpolating between these vacua. Collectively the $h^{\chi}$ describe the vevs of an adjoint Higgs field which break SU(3) to ${\rm U}(1) \times {\rm U}(1)$. Here they again form the boundary conditions for the domain wall.

In the pure SU(2) Yang-Mills theory domain walls form between vacua preserving different embeddings  of a ${\rm U}(1)_{\alpha}$ symmetry associated with magnetic charge \cite{Kobakhidze:2008zz}.

In both the above models there is an opportunity to trap gauge fields on the domain wall. This analysis can be generalized to SU(N) pure Yang-Mills theory where the rank of the algebra will again necessitate a systematic way of identifying all the boundary conditions for the domain walls.

\subsection{Vacuum alignment}\label{sec:Vacuumalignment}

Many extensions of the standard model feature multiple copies $\Phi_1,\ \Phi_2,\ \ldots$ of Higgs multiplets transforming according to a given representation of the gauge group $G$.  In general, their vevs $\langle \Phi_1 \rangle,\  \langle \Phi_2 \rangle,\ \ldots$ are not aligned in the internal representation space, so each multiplet breaks $G$ to a different subgroup, with the net unbroken symmetry being the intersection of all of these individual subgroups.  These subgroups may or may not be all isomorphic, depending on the model and the context.  For the cases where the individual subgroups are indeed isomorphic but differently embedded in the parent group $G$, then our analysis is relevant.

\section{Terminology and Notation}\label{sec:notation}

We now clearly outline some of the terminology and notation we use throughout this document. A reader who is familiar with standard notation in QCD and root systems may choose to skip this section and use it as a reference.

We choose to work exclusively with diagonal Cartan subalgebra generators, which can be done without loss of generality because given an arbitrary Cartan subalgebra it is always possible to simultaneously diagonalize each member using a similarity transformation within the Lie algebra. If the Lie algebra has rank $l$ we choose $ h^i \textrm{ where } i \in \{1,\dots,{\it  l}\}$ to refer to our basis for the Cartan subalgebra.

Throughout this document we physically contextualize our result using QCD and the weak force as examples. To do this we choose explicit representations. In each case we make use of the adjoint representation and the lowest dimensional fundamental representation, otherwise known as the smallest faithful representation.

In QCD for the 3 representation of SU(3) we use the Gell-Mann matrices $\lambda_1, \dots ,\lambda_8$ as generators. We refer to the gluons as a set of 8 Lorentz 4-vector fields $G^i_{\mu} \textrm{ where } i \in \{1,\dots, 8\}$ distributed over the Gell-Mann matrices; we write $X_i^{\mu} = G_i^{\mu} \lambda_i$ where there is no intended sum over $i$. We also make use of the linear combinations of the off diagonal gluons:
\begin{align}
Z^1_{\mu} = \frac{1}{\sqrt{2}} (G^1_{\mu} + i G^2_{\mu}),  && Z^2_{\mu} = \frac{1}{\sqrt{2}} (G^4_{\mu} +  i G^5_{\mu}),&&
Z^3_{\mu} = \frac{1}{\sqrt{2}} (G^6_{\mu} +  i  G^7_{\mu}).
\end{align}

\noindent Correspondingly we take linear combinations of the two diagonal gluons, renamed for notational convenience $G^3_{\mu} = A^1_{\mu} \textrm{ and } G^8_{\mu} = A^2_{\mu}$,
\begin{equation}
 B^p_{\mu} = A^i_{\mu} \alpha^p_i,
\end{equation}
\noindent where $p \in \{1,2,3\}$ and there is an implicit sum over $i$, which labels the components, $\alpha^p_i$, of the three roots $\alpha^1 = (1,0), \, \alpha^2 =(1/2, \sqrt{3}/2) , \, \alpha^3 = (-1/2, \sqrt{3}/2)$. In keeping with this notation we use a relabeling of the diagonal Gell-Mann matrices $\lambda_3  = A^1$ and $\lambda_8  = A^2$ to define the SU(3) Lie algebra generators $\kappa = A^i\alpha^1_i$, $\rho = A^i \alpha^2_i$ and $\epsilon = A^i \alpha^3_i$ associated respectively with $B^1_{\mu}, B^2_{\mu} \textrm{ and } B^3_{\mu}$. We also give rather unimaginative names to the Lie algebra generators associated with the valence gluons $Z^1_{\mu}, Z^2_{\mu} \textrm{ and } Z^3_{\mu}$:
\begin{align}
Z^1 = \lambda_1 + i \lambda_2, && Z^{-1} = \lambda_1 - i \lambda_2,\\
Z^2 = \lambda_4 + i \lambda_5, && Z^{-2} = \lambda_4 - i \lambda_5,\\
Z^3 = \lambda_6 + i \lambda_7, && Z^{-3} = \lambda_6 - i\lambda_7.
\end{align}

\noindent Notice these are precisely the raising and lowering operators of the SU(3) Lie algebra. Given a module $\phi = (\phi^1,\phi^2,\phi^3)$ for the fundamental representation of ${\rm SU}(3)$ these ladder operators can be used to raise (or lower) the states $\phi^p$ in this module which are associated with (can be directly labeled by) the weights:

\begin{equation}\label{eq:weightsofsu3}
u^1 = (\frac{1}{2}, \frac{1}{2\sqrt{3}}), \qquad u^2 =(-\frac{1}{2}, \frac{1}{2\sqrt{3}}) , \qquad u^3 = (0, -\frac{1}{\sqrt{3}}).
\end{equation}

Extending the SU(3) example we will refer to the I-spin, V-spin and U-spin directions in color space, which describe the three Cartan preserving embeddings of SU(2) inside SU(3). These are the three embeddings which have the Cartan subalgebra generators for SU(2) as a subset of the Cartan generators for SU(3). In terms of the Gell-Mann matrices, the generators of the SU(2) subgroup in each case are

\begin{eqnarray}\label{eq:embeddings}
\underbrace{\lambda_1, \lambda_2, \kappa}_{2\textrm{(I-spin)}}, \lambda_4, \lambda_5,\lambda_6, \lambda_7,\lambda_8&&\notag\\
\lambda_1, \lambda_2, \lambda_4, \lambda_5 ,\underbrace{\epsilon, \lambda_6, \lambda_7}_{2\textrm{(V-spin)}}, \epsilon' &&\notag\\
 \lambda_1 \lambda_2, \underbrace{\rho, \lambda_4, \lambda_5}_{2\textrm{(U-spin)}} ,\lambda_6, \lambda_7,\rho' &&
\end{eqnarray}
\noindent where we have chosen to introduce complementary matrices to the $\rho $ and $\epsilon$, namely $\rho' = -\sqrt{3}/2 A^1 + 1/2 A^2$ and $\epsilon' = \sqrt{3}/2 A^1 + 1/2 A^2$ respectively, so that each set of Lie algebra generators contains a diagonal Cartan subalgebra, which is orthogonal under the matrix trace.

In our weak force examples we use the Pauli matrices $\{\tau_1, \tau_2, \tau_3\}$  as a vector space basis for the adjoint representation (note $\tau_3$ is the diagonal generator of the weak isospin gauge group, $I_2$, in this representation) and the standard notation for the three gauge bosons $W^1_{\mu} = w^1_{\mu}\tau^1, W^2_{\mu} = w^2_{\mu}\tau^2, W^3_{\mu} = w^3_{\mu}\tau^3$ .

 Analogously to the QCD case we consider linear combinations of the weak force gauge bosons $W^+_{\mu} = w^+_{\mu}\tau^+ =  W^1_{\mu} - i W^2_{\mu},\, \,  W^{-}_{\mu} = w^-_{\mu}\tau^- = W^1_{\mu} + i W^2_{\mu} \textrm{ and } W^0_{\mu} = w^3_{\mu}\tau^3 = W^3_{\mu}$, and the corresponding generators $\tau^+ = \tau_1 - i \tau_2, \,\, \tau^- = \tau_1 + i \tau_2$ and $\tau_ 0~=~\tau_3$. We use this notation because +1, -1 and 0 are the respective ${\rm U}(1)_Q$ quantum numbers or electric charges of these linear combinations.

The adjoint action of the Lie algebra ${ \rm ad}_{h^i} \cdot E^{\alpha}$ on itself is defined by ${\rm ad}_{h^i}\cdot E^{\alpha} = [h^i, E^{\alpha}]$. In the special cases where the $E^{\alpha}$ are eigenvectors under the adjoint operation for some $h^i$ we write $ [h^i, E^{\alpha}] = \alpha(h^i) E^{\alpha}$

We say a linear transformation stabilizes a point if it maps that point back onto itself. For example if $\ket{\lambda}$ is an eigenvector of a Lie algebra generator $t_k \in {\cal L}$, so that $t^k \cdot \ket{\lambda} = \lambda(t^k) \ket{\lambda}$, then we say $\ket{\lambda}$ is stabilized by $t_k$.


\section{Root Systems, the Weyl group}\label{sec:weylgroups}

Our work relies heavily on the concept of roots and weights. Particle physicists often refer to the roots and weights as the quantum numbers of particles belonging to a nontrivial representation space of a non-Abelian gauge group. Consider the SU(2)-weak lepton doublet,
\begin{align}
{\it l}_L = \left(\begin{matrix} \nu_{e_L} \\ e_L\end{matrix}\right) \sim (1,2)(-1),
\end{align}
\noindent where by $(1,2)(-1)$ we mean the lepton doublet does not transform under the SU(3) color symmetry, however it transforms under a two dimensional representation of the SU(2) weak isospin gauge group and ${\it l} \rightarrow e^{-i \theta} {\it l}$ under the ${\rm U} (1)$ weak hypercharge symmetry. The SU(2) weights of the two states in this representation are the isospin quantum numbers of the fermions. The electron neutrino, $\nu_e$, has isospin quantum number $+1/2$. This is the highest weight of the representation. The electron, $e$, has isospin quantum number $-1/2$. This is the lowest weight of the representation.

The roots are the isospin charges of the self-interacting gauge bosons,
 \begin{equation}
  W^+_{\mu} =\left(\begin{matrix}  0 & w^+_{\mu} \\ 0 & 0\end{matrix}\right),  \qquad   W^-_{\mu} =\left(\begin{matrix} 0 & 0 \\  w^-_{\mu} & 0\end{matrix}\right).
  \end{equation}
\noindent The gauge bosons are associated with the SU(2) raising operator $\tau^+$, and the SU(2) lowering operator $\tau^-$ respectively. These are eigenstates of  the adjoint action of the weak-isospin generator $I_2$, that is ${\rm ad}_{I_2} \cdot \tau^{\pm} = [ I_2, \tau^{\pm}] = {\rm ad}_{I_2}(\tau^{\pm}) \tau^{\pm}$. The $+1$ isospin charge of $W^+$  and -1 isospin charge of $W^-$ follow from:
\begin{align}
\left[I_2, W^+_{\mu}\right] = w^+_{\mu} \left[I_2, \tau^+\right] = 1 W^+_{\mu} &&  \left[I_2, W^-_{\mu}\right] = w^-_{\mu} \left[I_2, \tau^-\right] = -1 W^-_{\mu}.\end{align}

\subsection{Constructing the Weyl group}
In general, it is possible to represent a semi-simple rank {\it l} Lie Algebra using two types of generators:
\begin{itemize}
\item{a set of {\it l} mutually commuting diagonalizable generators, $h^1, \dots, h^{\it l}$, which together with the linear combinations $\Sigma_i a^i h^i$, form a Cartan subalgebra, ${\cal C}_G$ and,}
 \item{a collection of simultaneous eigenstates $E^{\alpha}$ of the adjoint action of every Cartan subalgebra generator.}
\end{itemize}
\noindent Collectively the generators satisfy the commutation relations of the Lie algebra, ${\cal L}$,
\begin{equation}\label{eq:Liealgebradefinition}
\begin{gathered}\\
[h^i , h^j]  =  0 \\
 [h^i,E^{\alpha}]
 = {\rm ad}_{h_i} \cdot E^{\alpha} = \alpha(h^i) E^{\alpha}  \\
[E^{\alpha}, E^{-\alpha}] = h^{\alpha}\\
[E^{\alpha}, E^{\beta}] = N_{\alpha,\beta}E^{\alpha, \beta} \,\textrm{ if } \, \alpha \neq -\beta\\
\end{gathered}
 \end{equation}
\noindent where $h^{\alpha}$ is a linear combination of the $h^i$. We shall call $\alpha(h^i) = \alpha^i$ for convenience. Each eigenstate $E^{\alpha}$ can be labeled by an {\it l}-dimensional vector $\alpha = (\alpha^1,\dots,\alpha^{\it l})$ called a {\it root}. The root is a list of the {\it l} eigenvalues (structure constants) for the commutator, $[h^i , E^{\alpha}]$, of $E^{\alpha}$ with each $h^i \in {\cal C}_G$.

 The length of the roots depends on choosing a consistent normalization scheme for the generators. We fix the normalization of our Lie algebra generators by choosing ${\rm Tr }\, (E^{\alpha} E^{-\alpha}) = 2/(\alpha,\alpha)$, where $(a,b)$ is an invariant inner product\footnote{For example if one used an invariant inner product on the Lie algebra generators, such as the Cartan-Killing form or the regular trace and restricted this inner product to the Cartan generators then because the root space is dual to the Cartan subalgebra this induces an invariant inner product on the root space.}. This is a condition known as the Chevalley-Serre basis. It guarantees the components of the roots are integers.

 It follows from equation (\ref{eq:Liealgebradefinition}) that for each root $\alpha$, labeling a generator $E^{\alpha} \in {\cal L}$, there exists $-\alpha$, labeling the hermitian conjugate generator $E^{-\alpha} = E^{\alpha \dagger} \in {\cal L}$. We refer to $E^{\alpha}$ as a raising operator, and $E^{-\alpha}$ as a lowering operator. This leads us to partition the root system into two disjoint sets: the positive and the negative roots. We elect to call a root, $\alpha$, whose first nonzero component is positive, a ``positive root". The corresponding negated positive root, $-\alpha$, is termed a ''negative root". Not all of these roots are linearly independent. It is convenient to introduce a basis for the root space.

 A rank {\it l} Lie algebra has {\it l} independent Cartan subalgebra generators and therefore a set of {\it l} linearly independent  simple roots called  $\{\zeta^{(1)}, \dots,\zeta^{({\it l})}\}$. The simple roots are conventionally chosen to be the {\it l}-dimensional subset of the positive roots, with the property that every positive root can be written as a non-negative linear combination of $\{\zeta^{(1)}, \dots,\zeta^{({\it l})}\}$.

 It is clear from (\ref{eq:Liealgebradefinition}) that each root $\alpha$ is the pullback of a member of the Cartan subalgebra,
 \begin{equation}\label{eq:pullback}
 h^{\alpha} = \left[ E^{\alpha},E^{-\alpha}\right].
 \end{equation}
 Multiplying this expression on the left hand side by $h^j \in \{h^1,\dots, h^{\it l}\}$ and taking the matrix trace we see $h^{\alpha} =  \alpha^{\vee}_j h^j$ (sum over $ j \in \{1,\dots,{\it  l}\}$) where $\alpha^{\vee}_j = g_{ij} 2 \alpha^i/(\alpha, \alpha)$ (sum over $ i \in \{1,\dots, {\it l}\}$), where $g_{ij} = \left[{\rm Tr}\, (h^i h^j)\right]^{-1}$ is the inverse of the ${\it l}\times{\it l}$  matrix whose $ij$th element is $g^{ij} = \left[{\rm Tr}\, (h^i h^j)\right]$. We call $\alpha^{\vee} = 2\alpha/(\alpha,\alpha)$ a co-root; for example $\zeta^{(i)\vee} = 2\zeta^{(i)}/(\zeta^{(i)},\zeta^{(i)})$ is a simple co-root, for $\zeta^{(i)} \in \{\zeta^{(1)},\dots, \zeta^{\it (l)}\}$.

Linearity of the commutator bracket now allows us to extend our definition of the adjoint action to any $h^{\beta}$ acting on the Lie algebra according to
\begin{equation}\label{eq:adjoingtaction}
  {\rm ad}_{h^{\beta}} \cdot E^{\alpha}= \alpha(h^{\beta}) E^{\alpha} = (\alpha, \beta^{\vee}) E^{\alpha}.
  \end{equation}

\noindent The set of roots for a Lie algebra have the property that they completely characterize the Lie algebra. They also form a {\it crystallographic root system} $\Delta$ \cite{Kane:2001}, which is a set of roots with the property that $\forall \, \alpha, \beta, \gamma \in \Delta$:

\begin{enumerate}
\item{If $\alpha \in \Delta$, then $\chi \, \alpha \in \Delta$ if and only if $\chi = \pm 1$.}
\item{The reflection of $\beta$ in the hyperplane perpendicular to $\gamma$:
$s^{\gamma} \cdot \beta = \beta - (\beta, \gamma^{\vee}) \gamma $ also belongs to $\Delta$.\label{enumeratedweylatributes}}
 \item{$(\beta, \gamma^{\vee}) \in {\mathbb Z}$.}
    \end{enumerate}


Notice that condition (\ref{enumeratedweylatributes}) implies that $W = \{s^{\gamma} | \, \gamma \in \Delta\}$, the subset of symmetries of $\Delta$ generated by reflections in the hyperplanes orthogonal to the roots in $\Delta$, forms a reflection group known as the {\it Weyl group}.

 Any element of W can be expressed as a sequence of reflections in the simple roots.  This gives rise to a presentation of the Weyl group, called the Coxeter presentation, generated by reflections in the hyperplanes orthogonal to the simple roots, $\zeta^i$. If we refer to the angle between any two simple roots $\zeta^i$ and $\zeta^j$ as $\pi/ m_{ij}$, then the {\it Coxeter presentation} is:
\begin{equation}\label{eq:coxeterpresentation}
W = \left\{ s^{\zeta^i} | \left(s^{\zeta^i} s^{\zeta^j}\right)^{m_{ij}} = 1, \left(s^{\zeta^i}\right)^2  = 1 \right\}.
\end{equation}
\noindent The Coxeter presentation expression for each element, $w^{\gamma} \in W$, is not unique. However if we define the length of an expression to be the number of reflections, $s^{\zeta^i}$, it contains, then the relations can be used to reduce all Coxeter presentations for $w^{\gamma}$ to a fixed minimum length. This fixed length is a property of $\gamma$ relative to the choice of $\{\zeta^1,\dots,\zeta^{\it l}\}$.

To understand the relations in equation (\ref{eq:coxeterpresentation}) let ${\cal H}^{\zeta^{i\vee}}$ be the ({\it l} -1)-dimensional hyperplane orthogonal to $\zeta^i$. Because $\zeta^1$ and $\zeta^2$ are linearly independent, the intersection ${\cal H}^{\zeta^{1\vee}}\cap {\cal H}^{\zeta^{2\vee}}$ is an ({\it l} -2)-dimensional space, the complementary space being the plane spanned by $\zeta^1$ and $\zeta^2$. A reflection in ${\cal H}^{\zeta^{1\vee}}$ followed by a reflection in ${\cal H}^{\zeta^{2 \vee}}$ , $s^{\zeta^{1}}s^{\zeta^2}$, is the same as a rotation by twice the angle between ${\cal H}^{\zeta^{1\vee}}$ and ${\cal H}^{\zeta^{2\vee}}$ (that is a rotation by $2\pi/m_{12}$) in the $(\zeta^1,\zeta^2)$ plane. Therefore the relation $(s^{\zeta^1}s^{\zeta^2})^{m_{12}} = 1 $ is equivalent to the statement that $m_{12}$ concatenations of a rotation by an angle $2\pi/ m_{12}$ is the identity transformation.

The Weyl group has a natural analogue in the matrix picture \cite{Helgason}. Here conjugation by the operator

\begin{equation}\label{eq:weylgroupmatrices}
w^{\gamma} = {\rm exp}(E^{\gamma}) {\rm exp}(E^{-\gamma}) {\rm exp}(E^{\gamma}),
\end{equation}

\noindent acts on the Cartan subalgebra according to

\begin{equation}\label{eq:cartansubalgebraaction}
w^{\gamma} \cdot h^{\beta} = w^{\gamma} h^{\beta} w^{-\gamma} = (s^{\gamma}\cdot \beta^{\vee})_i h^i = (s^{\gamma} \cdot \beta)^{\vee}_i h^i,
\end{equation}

\noindent where $w^{-\gamma} =  \left({w^{\gamma}}\right)^{-1}$. We can check that (\ref{eq:weylgroupmatrices}) is a matrix representation for the Weyl group, acting as a module on the Cartan subalgebra ${\cal C}_G$, by checking that $w^{\gamma}\cdot h^{\beta} = h^{s^{\gamma} \cdot \beta}$. This follows directly from the action of $w^{\gamma} \cdot h^{\beta}$ on $E^{\alpha}$:
\begin{eqnarray}\label{eq:weyl1}
 [w^{\gamma} \cdot h^{ \beta}, E^{\alpha}] &=& (s^{\gamma} \cdot \beta)^{\vee}_i[h^i, E^{\alpha}] = \alpha^i (s^{\gamma}\cdot \beta)^{\vee}_i E^{\alpha} \notag\\ &= & (\alpha, (s^{\gamma}\cdot \beta)^{\vee})E^{\alpha} = [h^{s^{\gamma}\cdot \beta}, E^{\alpha}].
 \end{eqnarray}
\noindent Conversely conjugating (\ref{eq:weyl1}) by $w^{-\gamma}$:
\begin{eqnarray}\label{eq:weyl2}
 [h^{\beta}, w^{-\gamma} E^{\alpha} w^{\gamma}] &=& (s^{\gamma} \cdot \beta, \alpha^{\vee}) w^{-\gamma}E^{\alpha}w^{\gamma} \notag\\ &=& (\beta, (s^{\gamma} \cdot \alpha)^{\vee})w^{-\gamma} E^{\alpha} w^{\gamma} = [h^{\beta}, E^{s^{\gamma}\cdot \alpha}],
\end{eqnarray}
leads us to conclude $w^{-\gamma} \cdot E^{\alpha}  = E^{s^{\gamma} \cdot \alpha}$ and therefore (\ref{eq:cartansubalgebraaction}) also furnishes a matrix representation for the Weyl group acting as a module on the space of generators $\{E^{\alpha}| \, \,\alpha\in \Delta\}$. In the root system picture its elements are orthogonal transformations which act to permute the collection of roots belonging to $\Delta$.

In the matrix picture the Cartan subalgebra is an invariant subspace for the Weyl group and the Weyl group permutes the raising and lowering operators $E^{\alpha}$.

\subsection{Weights}\label{sec:weights}
More generally, the physical significance of being able to simultaneously diagonalize the Cartan subalgebra is that, for any representation space of the Lie group, there exists a basis, ${\cal B}$, of simultaneous eigenvectors, $\ket{u}$ , of the entire Cartan subalgebra.  Each eigenvector $\ket{u} \in {\cal B}$, can be labeled by the {\it l}-dimensional vector, $u = (u^1,\dots, u^{\it l}) = (u(h^1), \dots, u(h^{\it l}))$, formed by listing its eigenvalues, $h^i\ket{u} = u(h^i)\ket{u}$, for $h^i = h^1,\dots, h^{\it l}$. These {\it l}-dimensional vectors are the weights.

In the adjoint representation the weights are the root vectors. If the Lie group representation acts as a module over a vector space of $n$-dimensional column vectors (analogously to the SU(2)-weak lepton doublet), then the weights are the eigenvalues under left matrix multiplication by the Cartan subalgebra generators. The eigenvector labeled by the highest weight is annihilated by all raising operators.

The Weyl group action on the adjoint representation space eigenbasis, $w^{-\gamma}\cdot E^{\alpha}$ in equation (\ref{eq:weyl2}) (and on the weights of the adjoint representation, $s^{\gamma}\cdot \alpha$) can be generalized. The Weyl group reflection of a weight $u$ in the hyperplane orthogonal to root $\kappa$  is:
\begin{equation}\label{eq:weightaction}
s^{\kappa}\cdot u = u - (u,\kappa^{\vee})\kappa.
\end{equation}
In direct analogy to the adjoint representation, an arbitrary representation space for the Lie group furnishes a representation space for the Weyl group. This can be seen directly from the action of (\ref{eq:weylgroupmatrices}) on $\ket{u} \in {\cal B}$
\begin{equation}\label{eq:wrightbasisvectororbit} w^{-\kappa} \cdot \ket{u} = \ket{s^{\kappa}\cdot u}.\end{equation}
The result follows from analyzing the action of $h^i \in {\cal C}_G$ on $w^{-\kappa}\ket{u}$  which is described in full detail in Appendix A.

We introduce some terminology we need to talk about weights. The weights belonging to the Weyl group orbit of the highest weight are called extremal weights.

Consider a representation which has highest weight $\lambda$, and let $E^{\delta} \in {\cal L}$ be a generic raising operator for this representation. Then it is easy to see that each extremal weight $\mu = s^{\kappa} \cdot \lambda $ where $\kappa \in \Delta $, is also the highest weight with respect to a different choice of positive roots, as $\ket{\mu}$ is eliminated by an equivalent set of raising operators $w^{\kappa} \cdot E^{\delta} \in {\cal L}$. However the Weyl group permutes the set of Lie algebra generators, so both representations have the same generators. We would like to have a way of distinguishing between these representations and others which have qualitatively different sets of generators.

It is necessary to work with a basis for the weight space $\{\omega^1,\dots, \omega^{\it l}\}$ which is dual to the simple roots, that is $\omega^i \zeta^{j \vee} = \delta^{ij}$. We call $\{\omega^1,\dots, \omega^{\it l}\}$ fundamental weights. A linear combination of $\{\omega^1,\dots, \omega^{\it l}\}$ with non-negative coefficients is called a dominant weight. Every dominant weight is the highest weight of a representation, and up to conjugation by the Weyl group every highest weight is dominant.


\section{Lie subalgebras and embeddings}\label{sec:embeddings}

Lie subalgebras ${\cal L}_H \subset {\cal L}$ have generators labeled by closed subroot systems $\Delta_H \subset \Delta$, where by a {\it closed subroot system} \cite{Kane:2001} we mean

 \begin{enumerate}
\item{ A root system $\Delta_H \subset \Delta$, such that for all $\alpha, \beta \in \Delta_H$ if $\alpha + \beta \in \Delta$ then $\alpha + \beta \in \Delta_H$.}
\end{enumerate}

We can see ${\cal L}_H$ satisfies the Lie algebra commutation relations (\ref{eq:Liealgebradefinition}) because whenever  $E^{\alpha}, E^{\beta}\in {\cal L}_H$ and $N_{\alpha, \beta} \neq 0$, we have $[E^{\alpha}, E^{\beta}]\in {\cal L}_H$  (closure under the Lie bracket).

The Weyl group of the subroot system $\Delta_H$, $W_H = \{s^{\alpha} | \alpha \in \Delta_H\}$, is the subgroup of W, which permutes the subset of the roots belonging to $\Delta_H$ .

For each subroot system $\Delta_H$, or one of its Weyl group conjugates, there is a systematic way of choosing a basis of simple roots consisting of a proper subset $I_H \subset \{\zeta^1, \dots, \zeta^i\} \cup \{-\zeta_0\}$, of the union of the simple roots for $\Delta$ and, the negated highest weight of the adjoint representation (negated highest root). The method is given in the Borel-de-Siebenthal theorem (see Appendix B ). The Coxeter presentation for $W_H$ is generated by reflections in the hyperplanes $H^{\zeta^{j\vee}}, \, \zeta^j \in I_H$ orthogonal to the simple roots of $\Delta_H$.

The Weyl group action on the root system is regular (that is for all $\alpha, \beta \in \Delta$, there exists precisely one $s^{\gamma} \in W$ such that $\beta = s^{\gamma} \cdot \alpha$). The orbit $W \cdot \Delta_H$ represents all the embeddings of $\Delta_H$ inside $\Delta$. However we know that $W_{\Delta_H}$ maps $\Delta_H$ back onto itself, so each element in the orbit $W_{\Delta_H} \cdot \Delta_H = \Delta_H$ gives rise to the same embedding of ${\cal L}_H $ inside ${\cal L}$. Therefore the set $W/W_{\Delta_H} \cdot \Delta_H$ represents all the ``qualitatively different" embeddings of $\Delta_H$ and ${\cal L}_H$ inside $\Delta$ and ${\cal L}$ respectively. By ``qualitatively different" we mean the raising and lowering operators belonging to ${\cal L}_H$ and $w^{\kappa} {\cal L}_H w^{-\kappa}$ are distinct subsets of the full complement of raising and lowering operators belonging to ${\cal L}$.

One outcome of this is that we now know the number of embeddings of ${\cal L}_H$ inside ${\cal L}$ is $|W/W_{\Delta_H}|$. Because the Weyl group is finite we can simplify this expression\footnote{This follows from the orbit stabilizer theorem: Suppose that a linear algebraic group $G$ acts on the set X. If $G$ is finite then $|G| = |G\cdot x| \cdot |{\rm Stabilizer}(x)|$, that is, the order of the orbit of $x$, $|G\cdot x|$, divides $|G|$.}  to $|W|/|W_{\Delta_H}|$.

\section{Statement of Proof}\label{sec:proof}

 Here we write the vacuum manifold  $G/H$ in terms of adjoint Higgs vevs breaking $G  \rightarrow H_1 \rightarrow H_2 \rightarrow \dots \rightarrow H_{\it l}$, for some $H$ in this chain. We assume each of the embeddings $G  \supset H_1 \supset H_2 \supset \dots \supset H_{\it l}$ is Cartan preserving.

We have established that the embeddings of $H_1$ within $G$ arise from conjugation of the Lie algebra ${\cal L}_{H_1}$ for ${H_1}$ by the Weyl group $W/W_{\Delta_{H_1}}$ where $W_{\Delta_{H_1}}$ is the Weyl group of the maximal subgroup $H_1$. Moreover we know conjugation by any Weyl group element $w^{\kappa} \in W/W_{\Delta_{H_1}}$ acts on the Cartan subalgebra or vevs $h^1, \dots , h^{\it l}$ according to

\begin{equation}\label{eq:weylgroupactioncartansubalgebra}
w^{\kappa} \cdot h^j = \Sigma_i (\delta_{ij} - \Sigma_{n} \kappa^n \delta_{nj}\kappa^{\vee}_i) h^i = h^j - \kappa^j h^{\kappa}.
\end{equation}

\noindent So after identifying the generators(roots) excluded from the embedding of ${H_1}\subset G$ ($\Delta_{H_1} \subset \Delta$) we have a general formula for writing the vevs of the adjoint Higgs field $w^{\kappa}\cdot h^1, \dots, w^{\kappa} \cdot h^l$  causing the breaking of $G \supset w^{\kappa} {H_1} w^{-\kappa} \supset w^{\kappa} H_2 w^{-\kappa} \supset \dots \supset w^{\kappa} H_{\it l} w^{-\kappa}$ as linear combination of $h^1, \dots , h^{\it l}$ . If, after choosing an embedding of $H_1$ within $G$, identified with ${\cal L}_{H_1} \subset {\cal L}$, we wish to find all the different embeddings of $H_2$ within $H_1$ which have ${\cal L}_{H_2} \subset {\cal L}_{H_1}$ we simply repeat this procedure for $W_{\Delta_{H_1}}/W_{\Delta_{H_2}}$.

It is extremely simple to find $G/H$ when $H = H' \times {\rm U}(1)_H$ stabilizes (the representation space state labeled by) the highest weight of the lowest dimensional fundamental representation, $\ket{\lambda}$. For the adjoint representation we show that each vev in $G/H$ is $\Sigma_i \mu(h^i) h^i$  for an extremal weight $\mu$ of the fundamental representation, where $h_1,\dots, h_{\it l}$ break $G\rightarrow H_1, \dots, H_{\it l}$.

We first prove the adjoint Higgs vev, $h$, which breaks $G$ to $H$ is given by the linear combination $h =\Sigma_i \lambda(h^i)h^i$,  where the coefficients are the coordinates of highest weight of the fundamental representation. We then explain why other generators breaking $G$ to different embeddings $w^{\kappa} \cdot H$, $w^{\kappa} \cdot h = \Sigma_i \mu(h^i) h^i$,  are the linear combinations of $h^1,\dots, h^{\it l}$ which have the co-ordinates of the extremal weights, $\mu(h^i)$ as coefficients.

If $\Sigma_i \lambda(h^i)h^i$ is the adjoint Higgs vev which breaks $G$ to $H$, then it is the generator of the ${\rm U}(1)_H$ factor in $H = H' \times {\rm U}(1)_H$. Therefore  $\Sigma_i \lambda(h^i)h^i$ must stabilize $\ket{\lambda}$ (be a generator of $H$) and it must commute with each generator, $E^{\alpha} \in {\cal L}_G$, if and only if $E^{\alpha} \in {\cal L}_H$.

It is clear that $\Sigma_i \lambda(h^i) h^i$ is a generator of $H$ because

\begin{equation}
\Sigma_i \lambda(h^i) h^i \ket{\lambda} = \left(\Sigma_i \lambda(h^i)^2\right) \ket{\lambda}.
\end{equation}

 Furthermore, let $E^{\alpha} \in {\cal L}_H$. Then $E^{\alpha}$ is a raising or lowering operator and $E^{\alpha}$ stabilizes $\ket{\lambda}$, therefore we must have $E^{\alpha} \ket{\lambda} = 0$. If ${\alpha} \in \Delta_H$ then $-{\alpha} \in \Delta_H$, and by the same logic $E^{-\alpha}\ket{\lambda} = 0$. Consider the commutator
\begin{eqnarray}
[E^{\alpha}, \Sigma_i \lambda(h^i) h^i] &=& \Sigma_i \lambda(h^i) [E^{\alpha}, h^i] \notag\\
&=& \Sigma_i \lambda(h^i) \alpha^i E^{\alpha} \notag\\
& = &  \lambda(\Sigma_i\alpha^i h^i) E^{\alpha}\notag \\
& = & \lambda(h^{\alpha}) E^{\alpha}\notag\\
& = &  \lambda([E^{\alpha}, E^{-\alpha}]) E^{\alpha} \notag\\
& = & 0.
\end{eqnarray}
\noindent Therefore $\Sigma_i \lambda(h^i) h^i$ commutes with all the elements of ${\cal L}_H$.

Assume $\Sigma_i \lambda(h^i) h^i$ commutes with a generator $E^{\kappa} \not\in {\cal L}_{H}$ which does not belong to the Lie algebra of $H$, then we have
\begin{eqnarray}\label{eq:noncommute}
\Sigma_i \lambda(h^i) h^i &=& w^{\kappa} \Sigma_i \lambda(h^i) h^i w^{-\kappa}\notag\\
                          &=& \Sigma_i  \lambda(h^i) h^i - \Sigma_i \lambda(h^i)\kappa(h^i) h^{\kappa}\notag\\
                          &=& \Sigma_i  \lambda(h^i) h^i - \Sigma_{i j} \lambda(h^i)\kappa(h^i) \kappa^{\vee}(h^j) h^j\notag\\
                           &=& \Sigma_i  \lambda(h^i) h^i - \Sigma_{j} \frac{2 (\lambda, \kappa)}{(\kappa, \kappa)} \kappa(h^j) h^j\notag\\
                          & =&  \Sigma_i  \left(\lambda(h^i) - (\lambda,\kappa^{\vee})\kappa(h^i)\right) h^i \notag\\
                          & = &  \Sigma_i \left[ s^{\kappa} \cdot \lambda\right](h^i) h^i.
\end{eqnarray}
\noindent This creates a contradiction because we are insisting $E^{\kappa}$ does not stabilize $\ket{\lambda}$, so $w^{\kappa} \ket{\lambda} =\ket{s^{\kappa}\cdot \lambda} \neq\ket{\lambda}$ and the two sets of coefficients (of the linearly independent Cartan subalgebra generators $h^1,\dots, h^{\it l}$) in the above sum must be different. We have proved $\Sigma_i \lambda(h^i)h^i$ is the adjoint Higgs vev, $h$, which breaks $G$ to $H$.

Now each embedding $w^{\kappa} \cdot H = w^{\kappa} H w^{-\kappa}$ will stabilize a state in the representation labeled by an extremal weight $w^{\kappa} \cdot \ket{\lambda} = \ket{\mu}$. By the above argument, the center of the subgroup $w^{\kappa}\cdot H$ which stabilizes $\ket{\mu}$ is generated by $\Sigma_i \mu(h^i) h^i$. We have a remarkably easy formula for reproducing the vevs which break $G$ to all the different embeddings of the subgroup which stabilizes the highest weight of the lowest dimensional fundamental representation, $H$, as a linear combination of the Cartan subalgebra $h^1, \dots, h^{\it l}$.  Notice that $w^{\kappa}$ must belong to a nontrivial coset in $W/W_{\Delta_H}$, because conjugation by $w^{\kappa}$ only takes us from one embedding to another when $s^{\kappa}$ does not fix the highest weight.

We present a systematic method for determining the subgroup $H$ directly from the extended Dynkin diagram for the Lie group $G$. Each unmarked node in the extended Dynkin diagrams is labeled by a simple root. The node with a cross in the center is $\zeta^0$. To find the Dynkin diagram for $H$ we simply need to determine which of the simple roots in $\Delta$ are also in $\Delta_H$. We also need to work out if the highest root $\zeta^0$ is in $\Delta_H$. The subset of $\{\zeta^1,\dots, \zeta^{\it l}\} \cup \{-\zeta^0\}$ belonging to $\Delta_H$, will be the simple roots for $\Delta_H$.

First we determine which subset of the simple roots $\{\zeta^1,\dots,\zeta^{\it l}\}$ belong to $\Delta_H$. Take the highest weight, $\lambda$, and write it as a linear combination of the fundamental weights.
 \begin{equation}
 \lambda = a_1 \omega_1 + \dots + a_{\it l}\omega_{\it l}
 \end{equation}
We assume this highest weight is dominant ($a_1,\dots, a_{\it l} \ge 0$), if it is not then it is always possible to replace $\lambda$ by one of the extremal weights which is dominant. Construct a set $S_{\lambda} = \{ j |\, a_j =0\}$. For all $j \in S_{\lambda}$ we have $(\lambda, \zeta^{j\vee}) = 0$. We claim that $\zeta^j \in \Delta_H$, that is $E^{\pm \zeta^j}\ket{\lambda}  = 0 $, for all $j \in S_{\lambda}$. Otherwise if $E^{\pm \zeta^j}\ket{\lambda}  \neq 0 $ consider the norm $N_{\lambda \pm \zeta^j} =  \bra{\lambda} E^{\pm\zeta^j\dagger} E^{\pm\zeta^j}\ket{\lambda}$. Because $\lambda$ is the highest weight of the representation $N_{\lambda + \zeta^j} = 0$ while $ N_{\lambda - \zeta_j} = \bra{\lambda} [E^{\zeta^j}, E^{-\zeta^j}]\ket{\lambda} = \langle\lambda|\lambda\rangle (\lambda, \zeta^{j\vee}) = 0$. For the remaining simple roots labeled by $k \not \in S_{\lambda}$, we have $s^{\zeta^k}\cdot \lambda \neq \lambda$, therefore $w^{\zeta^k}\ket{\lambda} \neq \ket{\lambda}$ and from (\ref{eq:weylgroupmatrices}) we know that one of $E^{\pm \zeta^k}$ does not stabilize $\lambda$.

The highest root(negated highest root) $\pm\zeta^0$ does not belong to $\Delta_H$. This follows from the fact that $\zeta^0$ is some linear combination of all the simple roots (with positive coefficients), therefore if the set $S_{\lambda}$ is non-empty $(\lambda, \zeta^0) > 0$.


So the Dynkin diagram for $H$ can be reconstructed from the connected components of the Dynkin diagram for $G$ labeled by simple roots $\{\zeta^j | j \in S_{\lambda}\}$. This uniquely defines the non-Abelian factor $H'$ of $H$. The full subgroup $H$ which stabilizes the highest weight is a product of $H'$ with one  Abelian factor ${\rm U}(1)$ for each $k \not \in S_{\lambda}$. These extra ${\rm U}(1)$ factors are generated by the Cartan subalgebra generators $h^{\zeta^k}$, $k\not\in S_{\lambda}$, which (by definition) stabilize $\lambda$, even when the associated raising/lowering operators $E^{\zeta^k}$ do not.

If the Higgs field does not belong to the adjoint representation then the above analysis generalizes. The Weyl group reflections still give the different embeddings of the subgroup chain $G \supset H_1 \supset \dots \supset H_{\it l}$. If there is an associated Cartan subalgebra $h^1, \dots, h^{\it l}$ defined as the generators of ${\rm U}(1)_{H_i}$ factors appearing in the subgroup chain through $H_i = H_i' \times {\rm U}(1)_{H_1} \times \dots \times {\rm U}(1)_{H_i}$ (where $H_i'$ is some product of non-Abelian Lie groups) then equation (\ref{eq:weylgroupactioncartansubalgebra}) gives the linear combinations for the equivalent Cartan subalgebra generator for the ${\rm U}(1)_{w^{\kappa} \cdot H^i}$ factors belonging to the differently embedded subgroup chain $G \supset w^{\kappa} H_1 w^{-\kappa} \supset \dots \supset w^{\kappa} H_{\it l} w^{-\kappa} $, where $w^{\kappa} \in W/ W_{\Delta_H}$.

If a subgroup $ H \subset G$ annihilates a column vector $\ket{\nu}$, labeled by a weight $\nu$, then the differently embedded subgroup $w^{\kappa} H w^{-\kappa}$ annihilates the column vector $w^{\kappa} \ket{\nu}$. Hence if $\ket{\nu}$ breaks $G$ to $H$, then $\ket{s^{\kappa} \cdot \nu}$ breaks $G \supset w^{\kappa} H w^{-\kappa}$ and it follows directly from (\ref{eq:weightactionproof}) that equation (\ref{eq:weightaction}) gives the coordinates of the new weights as a linear combination of $\nu$ (and $\kappa$).

\section{Application of results to high-energy physics}
\label{sec:applyresults}

We wish to firmly ground the above discussion by applying the formulas from Sec.~\ref{sec:proof} to two explicit model building examples. We physically contextualize the key concepts in Secs.~\ref{sec:weylgroups} and \ref{sec:embeddings} via the smallest effective example: embeddings of U-spin, I-spin and V-spin within the SU(3) QCD gauge group. We also tackle the nontrivial problem of finding a full complement of  domain-wall boundary conditions for an adjoint Higgs field which break $\Gesix$ to different embeddings of $\Gsoten\times\Guone$, to demonstrate the effectiveness of the techniques developed in Sec.~\ref{sec:proof}.

\subsection{A quantum chromodynamics example}\label{sec:aquantumchromodynamicsexample}

Consider the Weyl group conjugations giving rise to differently embedded copies of the subgroups ${\rm SU}(2) \times {\rm U}(1)$ inside SU(3). Following \cite{Cho:2007ja} we rewrite the SU(3) pure Yang-Mills quantum chromodynamics Lagrangian in terms of the off diagonal gluons $Z^p_{\mu},\, p \in \{1,2,3\}$ and the dual potentials to the roots $B^p_{\mu}, \, p\in \{1,2,3\}$ defined in Sec. \ref{sec:notation}:

\begin{align}\label{lagrangian}
{\cal L} = - \frac{1}{4} {\cal G}_{\mu \nu} {\cal G}^{\mu\nu} = \sum_p \left\{ -\frac{1}{6}\left(F_{\mu \nu}^p\right)^2 + \frac{1}{2}\left|D_{p\mu} Z_{\nu}^{\,p} -D_{p\nu} Z^p_{\, \mu}\right|^2  - i g F^p_{\mu \nu}Z^{\mu\, *}_{\, p} Z^{\nu}_{\, p} \right. \notag \\ \left.- \frac{1}{2}g^2 \left[\left(Z^{p\,*}_{\, \mu} Z^p_{\,\nu}\right)^2 + \left(Z^{p\,*}_{\, \mu}\right)^2 \left(Z^p_{\nu}\right)^2\right]\right\}
\end{align}
\noindent where
\begin{align}
F^p_{\mu\nu} = \partial_{\mu} B^p_{\nu} - \partial_{\nu} B^p_{\mu}, && D_{p \mu} Z^p_{\nu} = \left(\partial_{\mu} - i g B^p_{\mu} \right)Z^p_{\nu}.
\end{align}

The Weyl group permutes the roots $\{\pm\alpha^1, \pm\alpha^2, \pm\alpha^3\}$ of the SU(3) Lie algebra. Hence the Weyl group action on the above Lagrangian will cause a permutation of the dual potentials $B^p_{\mu}, \, p \in \{1,2,3\}$. The orbit of each dual potential is described by the Weyl group action on the associated Cartan subalgebra element $A_i \alpha_i^p$. There will be a simultaneous permutation of the valence gluons $Z^p_{\mu}$ which correspond to the Weyl group orbits of the associated raising and lowering operators $Z^p, Z^{-p} , \, p \in \{1,2,3\}$. The permutation is concordant with the geometric picture of the Weyl group reflections of their root labels. Therefore the invariance of the above Lagrangian under Weyl group reflections is encapsulated in the sum over the index $p$.

The clarity of this presentation is a direct consequence of the associated generators $\epsilon, \rho $ and  $Z^{\pm p}\,\textrm{ where } p \in \{1,2,3\}$ (it is not necessary to include $\kappa$ in this list, because SU(3) has rank 2, however we can substitute it for either $\epsilon$ or $\rho$ if we wish) forming a useful computational basis for the Lie algebra: the Chevalley basis. Here each of the three subset $\{\kappa, Z^{\pm 1}\}$, $ \{\rho, Z^{\pm 2}\}$ and $\{\epsilon, Z^{\pm 3}\}$ defines an embedding of SU(2) inside SU(3). These correspond to the closed crystallographic root systems $\{\pm \alpha^1\}$ whose Weyl group fixes a point on the hyperplane orthogonal to $\alpha^1$, $\{\pm \alpha^2\}$ and $ \{\pm \alpha^3\}$, whose Weyl groups fix analogous points. Cross checking this with Sec. \ref{sec:notation} we see these are precisely the I-spin, V-spin and U-spin embeddings. Each embedding commutes with one of the Abelian subgroup generators $\lambda_8 (= \kappa'), \rho' \textrm{ or } \epsilon'$ which we now have the tools to write as  $\Sigma_i \mu(h^i) h^i$ for any diagonal Cartan subalgebra $\{ h^1, h^2\}$ for SU(3) (in Sec. \ref{sec:notation} our Cartan subalgebra was chosen to be $\lambda_3$ and $\lambda_8$) and the three extremal weights of the lowest dimensional fundamental representation for SU(3).

We can adapt this scenario to incorporate interactions between quarks and ${\rm SU}(3)$ gauge fields. The quarks belong to a $3$-dimensional module, $Q$, for the fundamental representation of ${\rm SU}(3)$. According to subsection \ref{sec:weights} we can use the weights of the fundamental representation $\vec{u}^p$ to label a basis $|u^p\rangle, \, p \in \{1,2,3\}$  for the module $Q$. With respect to this basis we write the quark field components as $q^p,$ for $p =1,2,3$ (for the wights given in equation (\ref{eq:weightsofsu3}) this will lead to the interpretation that the quarks field components are precisely the standard $r,g,b$). We illustrate the specific example of quarks interacting with an Abelianized gauge potential which mediates interactions between quarks carrying the same colour charge. This example can be used to study quark-gluon plasmas \cite{Cho:2007ja};  using the Lagrangian
\begin{equation}\label{eq:quarkgluonplasma}
{\cal L}_{quarks} = -\frac{1}{8}\sum_p  \left(\tilde{F}_{\mu \nu}^p\right)^2 + \sum_p \overline{q}^p \left(i\gamma^{\mu}\tilde{D}_{p\mu} - m\right) q^p,
\end{equation}
where in terms of the Abelian guage fields\footnote{This redefinition of the diagonal gluons introduces superficial differences between equations (\ref{eq:quarkgluonplasma}) and (\ref{lagrangian}). In equation (\ref{lagrangian}) we use the dual potentials to the roots $B^p_{\mu}$ to define the diagonal gluons since the coupling terms between valence gluons and dual potentials are extremely simple. This happened because the valence gluons directly correspond to raising/lowering operators of the ${\rm SU}(3)$ Lie algebra. In equation (\ref{lagrangian}) writing the Cartan generators as $B^p = A^i \alpha_i^p$ while using $Z^{\pm p}$ for the raising/lowering operators implies we are directly choosing to define our Cartan subalgebra through equation (\ref{eq:pullback}), giving us access to the relations in equation (\ref{eq:adjoingtaction}). This is particularly helpful for the pure Yang-Mills gauge theory Lagrangian. In equation (\ref{eq:quarkgluonplasma}) we simplify the quark gluon vertex terms by define our Abelian gauge fields in terms of the weights. At the end of the day it doesn't matter how we write the diagonal gluons. Our choice reflects the easiest way to introduce coupling between these diagonal gluons and other fields in the Lagrangian. This is in part why we choose to work with quarks interacting through an Abelianized  gauge potential.} $A^p_{\mu} = A^i_{\mu} u_i^p$,
\begin{align}
\tilde{F}^p_{\mu\nu} = \partial_{\mu} A^p_{\nu} - \partial_{\nu} A^p_{\mu}, && \tilde{D}_{p\mu} q^p = \left(\partial_{\mu} - i g A^p_{\mu}\right)q^p.
\end{align}
The gauge field kinetic term is equivalent to the Abelian field strength tensor in equation (\ref{lagrangian}). The second term in equation (\ref{eq:quarkgluonplasma}) describes the prorogation and interactions of quarks with the Abelian gauge fields. Because the Weyl group action permutes the weights of the fundamental representation $\vec{u}^p$ it also causes a a permutation of the Abelian gauge fields $A^p_{\mu}, \, p \in \{1,2,3\}$ which match the orbits of the corresponding Cartan subalgebra generators $A_i u^p_i$. This permutation simultaneously acts on the quark field components, $q^p$, which undergo an identical orbit due to the Weyl group action on the basis vectors $|u^p \rangle$, see equation (\ref{eq:weightactionproof}) in the Appendix. The summation over $p$ ensures equation (\ref{eq:quarkgluonplasma}) is always invariant under Weyl group permutations.

\subsection{A non-adjoint ${\rm SU}(3)$ Higgs field}

For an example of a non-adjoint Higgs, we also examine a triplet $\phi = (\phi^1,\phi^2,\phi^3)$ of Higgs fields transforming as a module under the fundamental representation of a global ${\rm SU}(3)$ symmetry. We identify each Higgs field component with one of the weights of ${\rm SU}(3)$ given in equation (\ref{eq:weightsofsu3}). Consider the Higgs potential
\begin{equation}
V = -\frac{1}{2} \mu^2
\phi^{\dagger}\phi  + \frac{1}{4} \lambda\left(\phi^{\dagger}\phi\right)^2,
\end{equation}
where $\mu^2,\lambda \ge 0$ are free parameters. The minimum of this potential occurs for nonzero values of $|\phi|^2$, under these circumstances the third component of the Higgs triplet can develop a nonzero vacuum expectation value,
\begin{equation}
\langle \phi^i \rangle = v \delta_{i3} .
\end{equation}
This vacuum breaks the global ${\rm SU}(3)$ symmetry down to the I-spin embedding of ${\rm SU}(2)$ referred to in subsections \ref{sec:aquantumchromodynamicsexample} and \ref{sec:notation}. We identify the vacuum expectation value with the root $u^3 = (0, -1/\sqrt{3})$ from equation (\ref{eq:weightsofsu3}). The weights labeling other elements of the vacuum manifold can be generated by the Weyl group orbit of $u^3$. We can explicitly calculate the other elements of the vacuum manifold using equation (\ref{eq:weightaction}) in conjunction with $(u^3, \alpha^{p\vee}) = -\delta_{3p} -\delta_{29}$ (for the roots and weights defined in Sec. \ref{sec:notation}). This gives:
\begin{eqnarray}
s^{\alpha^1} \cdot u^3 &=& u^3, \notag\\
s^{\alpha^2} \cdot u^3 &=& u^3 + \alpha^2 = u^1,\notag\\
s^{\alpha^3} \cdot w^3 &=& u^3 + \alpha^3 = u^2,
\end{eqnarray}
where the properties of the ladder operators $Z^p$ can be conveniently used to identify a simplified expression. Colloquially, we say that the ladder operators,
raise (lower) states in the fundamental representation because $Z^p|u^j\rangle \propto | \alpha^p + u^j\rangle$, where the latter corresponds to a higher (respectively lower) weight of the representation. This is easily verified through the commutation relations (\ref{eq:Liealgebradefinition}), of the ladder operators $[Z^p, h^i] = \alpha^p(h^i) Z^p$ with the Cartan subalgebra generators $h^i$. The three different vacua associated with the above weights: $\langle \phi^i \rangle = v \delta_{ik}$ for $k = 3, 1$ and $2$ respectively, break ${\rm SU}(3)$ to the I-spin, V-spin and U-spin embeddings of ${\rm SU}(2)$.

\subsection{Adjoint Higgs field domain-wall-brane boundary conditions breaking $\Gesix\to\Gsoten\times\Guone$}
We use the method developed in the previous section to find all the adjoint Higgs vevs which break $\Gesix$ to all the different embeddings of $\Gsoten\times\Guone$; this example is directly motivated by an extra-dimensional ``clash of symmetries" domain-wall brane model \cite{Davidson:2007cf}. Our choice of Cartan subalgebra for $\Gesix$ is given in Table~\ref{tab:e6:gen}; the entries of this table follow directly from the branching rules \cite{Slansky:1981yr}:
\begin{eqnarray}
\Gesix &\supset&  \Gsoten \times {\rm U}(1)_{h^1} \supset {\rm SU}(5) \times {\rm U}(1)_{h^1} \times {\rm U}(1)_{h^2}\notag\\
& \supset& {\rm SU}(3) \times {\rm SU}(2) \times {\rm U}(1)_{h^1} \times {\rm U}(1)_{h^2} \times {\rm U}(1)_{h^3}\notag\\
& \supset& {\rm SU}(3) \times {\rm U}(1)_{h^1} \times {\rm U}(1)_{h^2} \times {\rm U}(1)_{h^3} \times {\rm U}(1)_{h^4}\notag\\
& \supset& {\rm SU}(2)  \times {\rm U}(1)_{h^1} \times {\rm U}(1)_{h^2}\times {\rm U}(1)_{h^3} \times{\rm U}(1)_{h^4}\times {\rm U}(1)_{h^5}\notag\\
& \supset& {\rm U}(1)_{h^1} \times {\rm U}(1)_{h^2} \times {\rm U}(1)_{h^3} \times {\rm U}(1)_{h^4}\times {\rm U}(1)_{h^5}\times {\rm U}(1)_{h^6}.\label{eq:E6 branching}
\end{eqnarray}

\begin{table}[htpb]
\centering
\begin{tabular}{|r|rrrrrr|}
\hline
     & $60h^1$ & $60h^2$ & $60h^3$ & $60h^4$ & $60h^5$ & $60h^6$ \\
\hline
\hline
1  & $20$ & $0$ & $0$ & $0$ & $0$ & $0$ \\
2  & $-10$ & $2\sqrt{15}$ & $3\sqrt{10}$ & $-5\sqrt{6}$ & $0$ & $0$ \\
3  & $-10$ & $2\sqrt{15}$ & $3\sqrt{10}$ & $5\sqrt{6}$ & $0$ & $0$ \\
4  & $-10$ & $2\sqrt{15}$ & $-2\sqrt{10}$ & $0$ & $5\sqrt{2}$ & $5\sqrt{6}$ \\
5  & $-10$ & $2\sqrt{15}$ & $-2\sqrt{10}$ & $0$ & $5\sqrt{2}$ & $-5\sqrt{6}$ \\
6  & $-10$ & $2\sqrt{15}$ & $-2\sqrt{10}$ & $0$ & $-10\sqrt{2}$ & $0$ \\
7  & $-10$ & $-2\sqrt{15}$ & $-3\sqrt{10}$ & $5\sqrt{6}$ & $0$ & $0$ \\
8  & $-10$ & $-2\sqrt{15}$ & $-3\sqrt{10}$ & $-5\sqrt{6}$ & $0$ & $0$ \\
9  & $-10$ & $-2\sqrt{15}$ & $2\sqrt{10}$ & $0$ & $-5\sqrt{2}$ & $-5\sqrt{6}$ \\
10 & $-10$ & $-2\sqrt{15}$ & $2\sqrt{10}$ & $0$ & $-5\sqrt{2}$ & $5\sqrt{6}$ \\
11 & $-10$ & $-2\sqrt{15}$ & $2\sqrt{10}$ & $0$ & $10\sqrt{2}$ & $0$ \\
12 & $5$ & $-5\sqrt{15}$ & $0$ & $0$ & $0$ & $0$ \\
13 & $5$ & $3\sqrt{15}$ & $-3\sqrt{10}$ & $5\sqrt{6}$ & $0$ & $0$ \\
14 & $5$ & $3\sqrt{15}$ & $-3\sqrt{10}$ & $-5\sqrt{6}$ & $0$ & $0$ \\
15 & $5$ & $3\sqrt{15}$ & $2\sqrt{10}$ & $0$ & $-5\sqrt{2}$ & $-5\sqrt{6}$ \\
16 & $5$ & $3\sqrt{15}$ & $2\sqrt{10}$ & $0$ & $-5\sqrt{2}$ & $5\sqrt{6}$ \\
17 & $5$ & $3\sqrt{15}$ & $2\sqrt{10}$ & $0$ & $10\sqrt{2}$ & $0$ \\
18 & $5$ & $-\sqrt{15}$ & $\sqrt{10}$ & $-5\sqrt{6}$ & $5\sqrt{2}$ & $5\sqrt{6}$ \\
19 & $5$ & $-\sqrt{15}$ & $\sqrt{10}$ & $-5\sqrt{6}$ & $5\sqrt{2}$ & $-5\sqrt{6}$ \\
20 & $5$ & $-\sqrt{15}$ & $\sqrt{10}$ & $-5\sqrt{6}$ & $-10\sqrt{2}$ & $0$ \\
21 & $5$ & $-\sqrt{15}$ & $\sqrt{10}$ & $5\sqrt{6}$ & $5\sqrt{2}$ & $5\sqrt{6}$ \\
22 & $5$ & $-\sqrt{15}$ & $\sqrt{10}$ & $5\sqrt{6}$ & $5\sqrt{2}$ & $-5\sqrt{6}$ \\
23 & $5$ & $-\sqrt{15}$ & $\sqrt{10}$ & $5\sqrt{6}$ & $-10\sqrt{2}$ & $0$ \\
24 & $5$ & $-\sqrt{15}$ & $-4\sqrt{10}$ & $0$ & $-5\sqrt{2}$ & $-5\sqrt{6}$ \\
25 & $5$ & $-\sqrt{15}$ & $-4\sqrt{10}$ & $0$ & $-5\sqrt{2}$ & $5\sqrt{6}$ \\
26 & $5$ & $-\sqrt{15}$ & $-4\sqrt{10}$ & $0$ & $10\sqrt{2}$ & $0$ \\
27 & $5$ & $-\sqrt{15}$ & $6\sqrt{10}$ & $0$ & $0$ & $0$ \\
\hline
\end{tabular}
\caption{
\small
The six diagonal generators $h^\text{1--6}$ of $\Gesix$.
The diagonal elements of the generator $h^n$ are found by taking
the $n^\text{th}$ column and multiplying it by $1/60$. Also the
rows give the coefficients $f_\text{1--6}$ of these generators that yield a
linear combination that breaks $\Gesix\to\Gsoten\times\Guone$.}
\label{tab:e6:gen}
\end{table}

\begin{figure}[htpb]
\centering
\includegraphics[width=\textwidth]{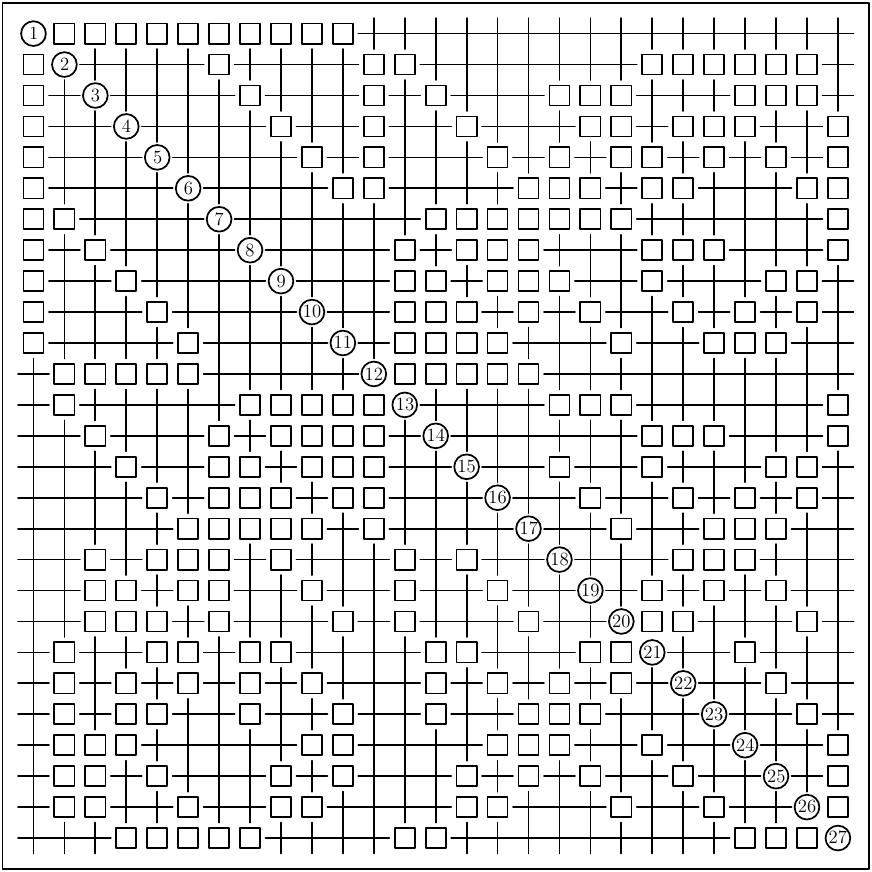}
\caption{
\small
A pictorial representation of the twenty-seven rearrangements of
the diagonal generator $h^1$ of $\Gesix$.
Each rearrangement can be reconstructed from one of the twenty-%
seven rows (or columns) of symbols in this picture.  To find the
diagonal entries of the $n^\text{th}$ rearrangement, read along
the $n^\text{th}$ row and translate the symbols according to:
circles $\bigcirc$ correspond to the single $1/3$ entry, squares
$\Box$ to $-1/6$ and crosses $+$ to $1/12$ (note that adjacent
crosses are touching).  The number in the center of each circle
tells its row and column number (being the same).  Row $n$ of
this picture corresponds precisely to row $n$ of Table~%
\ref{tab:e6:gen} in the sense that the linear combination
$\sum_{a=1}^6f_ah^a$, where the $f_\text{1--6}$ are chosen from
row $n$ of Table~\ref{tab:e6:gen}, yields the rearranged version
of the generator $h^1$ represented by the symbols of row $n$ in
this picture.
}
\label{fig:e6:epretty}
\end{figure}

 As mentioned in Sec.~\ref{sec:motivation} our primary motivation for studying this problem arose from a co-dimension-1 clash-of-symmetries domain-wall brane. The brane originates from an $\Gesix$ adjoint Higgs field ${\cal X}$ which condenses spontaneously to break translational invariance along the extra dimension of a 4+1-dimensional space-time manifold.

 The Lagrangian for this theory is invariant under a $Z_2 \times \Gesix$ internal symmetry. It is a linear combination of the invariant kinetic term ${\rm Tr} \left[D^{\mu} {\cal X} D_{\mu}{\cal X}\right]$ and a potential
 formed from the $\Gesix$ Casimir invariants $I_2 = {\rm Tr} {\cal X}^2$ and $I_6 = {\rm Tr} {\cal X}^6$ as well as the powers $I_2^2$ and $I_2^3$. Casimir invariants corresponding to odd powers of ${\cal X}$ must be omitted due to the imposed $Z_2$, ${\cal X} \rightarrow -{\cal X}$, symmetry. The potential is truncated at 6th order because the coupling constants of higher order invariants have negative mass dimensions and are therefore suppressed by powers of the putative ultraviolet completion scale (see Sec.~\ref{sec:motivation}), yet the  fourth order invariants exhibit an accidental $O(78)$ symmetry, so we must include a ${\rm Tr}{\cal X}^6$ term. A subset of the local minima of the Casimir invariants occur at adjoint Higgs vevs which break $Z_2 \times \Gesix \rightarrow \Gsoten \times \Guone$. If the solution ${\cal X}$ to the associated Euler-Lagrange equations interpolates between vacuum expectation values which break $Z_2 \times \Gesix$ to a specific pair of differently embedded copies of $\Gsoten \times \Guone$ then \cite{Davidson:2007cf} postulates a copy of the standard model particles can be trapped on the 3+1-dimensional domain-wall brane. To find ${\cal X}$ it is necessary to write the boundary conditions at the two antipodal extremes of the extra dimension as a linear combination of the adjoint Higgs vevs $h^1,\dots , h^6$, the generators of the Abelian subgroup factors given in Eq.~\eqref{eq:E6 branching}.

Because $\Gsoten \times \Guone$ stabilizes the highest weight of the lowest dimensional fundamental representation for $\Gesix$ this is now a trivial problem. Each of the possible boundary conditions which break $\Gesix \to \Gsoten \times \Guone$ can be written as a linear combination of $h^1,\dots, h^6$ using $\Sigma_i \mu(h^i) h^i$ where $\mu$ is one of the 27 extremal weights of the lowest dimensional fundamental representation for $\Gesix$. Explicitly the 27 different vevs breaking $\Gesix \rightarrow \Gsoten \times \Guone$ are
\begin{equation}
    \langle {\cal X} \rangle \propto \Sigma_{a=1}^{6} f_a h^a
\end{equation}
where the sextuplet $f_{1,\dots , 6}$ takes values from one of the rows of the Table~\ref{tab:e6:gen}. In Fig.~\ref{fig:e6:epretty} we have reproduced a figure from Ref.~\cite{DPGeorge2009} which graphically identifies the diagonal entries of each of these 27 vevs breaking $\Gesix\rightarrow\Gsoten\times\Guone$.

\section{Conclusion}
\label{sec:conclusion}

    We have shown how to write the vacuum manifold $G/H$ as a linear combination of vevs breaking $ G \supset H_1  \supset H_2  \supset \dots \supset H_{\it l}$, for the case where the embedding of $H \subset G$ is Cartan preserving and $H$ is a subgroup in the above chain. We have highlighted the simple case when the Higgs field is in the adjoint representation and $H$ stabilizes the highest weight of the lowest dimensional fundamental representation for $G$ and complemented our discussion with physical examples.


    Our work is motivated by current research in high-energy physics, where symmetry breaking is used extensively, and where an explicit and exhaustive construction of embeddings is of direct relevance.  In subsections \ref{sec:flippedgrandunified}-\ref{sec:Vacuumalignment} we have presented four physical contexts where our results can be directly applied, and in two of these four model building scenarios we apply our results to the contemporary research papers \cite{Cho:2007ja} and \cite{Davidson:2007cf}.

\begin{acknowledgments}

This work was supported in part by the Australian Research Council and in part by the Puzey Bequest to the University of Melbourne. DG was supported by the Netherlands Foundation for Fundamental Research of Matter (FOM), and the Netherlands Organisation for Scientific Research (NWO).

\end{acknowledgments}


\section{Appendix A} \label{sec:appendixA}
We explicitly evaluate the action of $h^i$ on  $w^{-\kappa} \ket{\nu}$:

\begin{eqnarray}\label{eq:weightactionproof}
h^i w^{-\kappa} \ket{\nu} &=&  w^{-\kappa} \left(w^{\kappa} h^i w^{-\kappa}\right) \ket{\nu} \notag \\
&=&  w^{-\kappa} \left(h^i - \kappa^i h^{\kappa}\right) \ket{\nu} \notag\\
&=&  w^{-\kappa} \left(h^i - \kappa^i \kappa_j^{\vee}h^j\right) \ket{\nu}\notag\\
&=& \left(\nu^i -  (\nu, \kappa^{\vee})\kappa^i\right)  w^{-\kappa} \ket{\nu} \notag\\
&=& \left[ s^{\kappa}\cdot \nu \right]^i  w^{-\kappa} \ket{\nu}\notag\\
& =& h^i\ket{s^{\kappa}\cdot \nu}.
\end{eqnarray}
where  to get the second equality we have used $h^j = \Sigma_i \delta_{i}^j h^i$ in Eq.~ \eqref{eq:cartansubalgebraaction}. Thus because of Eqs.~\eqref{eq:weylgroupmatrices} and~\eqref{eq:cartansubalgebraaction} it is not a coincidence that every representation space for the Lie group furnishes a representation space for the Weyl group.

\section{Appendix B}\label{sec:appendix}

The Borel-de-Siebenthal theorem gives a systematic way of identifying the maximal Lie subgroups of $G$ directly from the extended Dynkin diagram for $G$ \cite{Kane:2001}. It does this by identifying which subset of the simple roots belonging to $\Delta$, the root system for the Lie algebra ${\cal L}$ of $G$, are also simple roots for $\Delta_H$, the subroot system of the Lie algebra ${\cal L}_H$ of maximal subgroup $H \subset G$. The nodes of the extended Dynkin diagram of $G$ which are labeled by simple roots $\zeta^i$ which {\bf do not} belong to $\Delta_H$ are then removed, along with all their adjacent edges. The remaining graph is the Dynkin diagram for $H$. If the subgroup $H \subset K \subset G$, is not maximal then this procedure can be iterated to determine $K\subset G$ and $H\subset K$.

\begin{theorem}[Borel-de-Siebenthal] Let $\Delta$ be an irreducible crystallographic root system. Let $\{\zeta^1,\dots, \zeta^{\it l}\}$ be the simple roots for $\Delta$. Let $\zeta^0$ be the highest root of $\Delta$ with respect to $\{\zeta^1,\dots, \zeta^{\it l}\}$. Expand:

\begin{equation}
\zeta^0 = \Sigma_i c_i \zeta^i
\end{equation}

\noindent Then the maximal closed subroot systems of $\Delta$ (up to Weyl group reflections) are those with fundamental systems

\begin{itemize}
\item{$\{\zeta^1,\zeta^2,\dots,\hat{\zeta}^i,\dots,\zeta^{\it l}\}$ where $c_i = 1$;}
\item{$\{-\zeta^0,\zeta^1,\dots,\hat{\zeta}^i,\dots , \zeta^{\it l}\}$ where $c_i = p$ (prime)}
\end{itemize}
\noindent Where ``$\hat{\zeta}^i$" is being used to denote elimination.
\end{theorem}

\end{document}